\documentclass[aps, showpacs,pra,twocolumn] {revtex4}
\usepackage{hyperref}
\usepackage{amsmath }
\usepackage{amssymb}
\usepackage{epsfig}
\usepackage{natbib}

\begin{document}
\bibliographystyle{revtex}

\title{ Functional minimization method addressed to the vacuum finding
 \\
  for an arbitrary driven quantum oscillator }

\author{S.V. Anischenko}
\email{Lesavik@yandex.ru}
 \affiliation{Belarus State University, Bobruiskaya 5, Minsk 220050,
Belarus}
\author{S.L. Cherkas}
\email{cherkas@inp.minsk.by}
 \affiliation{Institute for Nuclear Problems, Bobruiskaya
11, Minsk 220050, Belarus}
\author{V.L. Kalashnikov}
\email{v.kalashnikov@tuwien.ac.at}
 \affiliation{Institut f\"{u}r Photonik, Technische
Universit\"{a}t Wien, Gusshausstrasse 27/387, Vienna A-1040,
Austria}

\received{ \today }

\begin{abstract}The old problem exists for a driven
(time-dependent) quantum oscillator: to differ the true vacuum
state from the squeezed one. We suggest finding the true vacuum
state by minimization of the functional containing the difference
of the potential and kinetic energies of oscillator. Analytical
and numerical examples confirming this offer are considered.
\end{abstract}
\maketitle

\section{Introduction}

A time-dependent (driven) oscillator arises naturally in a number
of fields of the theoretical physics \cite{dek,birrel}. In
particular, it has an application in cosmology and astrophysics,
where the scalar, fermion, gravitational, and other quantum fields
evolve in an expanding Universe \cite{birrel,lin}. Particle
creation by the nonstationary gravitational field is long
considered as one of the possible sources of the matter origin in
the Universe, and, to talk about a ``particle'' one has to
understand what is the vacuum. It should also be mentioned that
according to the modern view the vacuum fluctuations  were the
seeds for the structure formation in Universe \cite{liddle}.

Nevertheless, the definition of the ground (vacuum) state remains
to be obscure \cite{birrel,park,kim,proc}. This forces one to use
the well-known adiabatic states in concrete calculations
\cite{win}, whereas for systems which admit analytical
consideration, for instance, quantum field in the De Sitter
Universe, Bunch-Davis vacuum states \cite{banch} can be built.
However it would be desirable to define vacuum state without
appealing to the adiabatic series or analytical solution (it may
be impossible).
 This issue is addressed in our paper. The suggested
method allows finding numerically the true vacuum state (if it
exists).

Let us remind the problem in more detail.

Hamiltonian of the time-dependent oscillator has the following
form:
\begin{equation}
H=\frac{1}{2}{\dot x}^{2}+\frac{1}{2}\omega^2(t) x^2.
\end{equation}

Its quantization in the Heisenberg picture consists in a
replacement of the coordinate $x$ by the time-dependent operator:
\begin{equation}
{\hat x}(t)={\hat {\mbox{a}}}\, u(t)+ {\hat{\mbox{a}}}^+ u^*(t),
\end{equation}
where the operators ${\hat {\mbox{a}}}$ and ${\hat {\mbox{a}}}^+$
obey the commutator relation
\begin{equation}
[{\hat {\mbox{a}}},{\hat {\mbox{a}}}^+]=1,
\end{equation}
whereas the function $u$ satisfies the equation
\begin{equation}
{\dot u}^{*}u-\dot u\, u^*=\mbox{i}. \label{rel0}
\end{equation}
These rules provide the standard commutation relation for the
momentum and coordinate operators
\begin{equation}
[\hat p,\hat x]=[{\hat {\dot x}},\hat x]=-\mbox{i}.
\end{equation}
The vacuum state is defined as the state, which is the null-space
of the annihilation operator: $\hat a|0>=0$. However, there
remains a problem in the concrete definition of $u$. This function
satisfies the oscillator equation of motion
\begin{equation}\ddot u+\omega^2 u=0, \label{osc}
\end{equation}
 and one has to
define (at some instant) the initial condition corresponding to
the true vacuum state. It should be noted that there exists a
family \cite{birrel,kim} of the functions $u$, which satisfy Eq.
(\ref{rel0}) and are interrelated by the Bogolubov's
transformation:
\begin{eqnarray}
u(t)=\cosh r \,u_0(t)+\sinh r\, \mbox{e}^{\mbox{i} \delta} u_0^*(t)\nonumber\\
u^*(t)=\cosh r\, u_0^*(t)+\sinh r\, \mbox{e}^{-\mbox{i} \delta}
u_0(t). \label{bog}
\end{eqnarray}

In the terms of an oscillator with constant frequency
$\omega=const$, this means that it is necessary to differ the true
vacuum from the squeezed vacuum states. When $\omega=const$ the
vacuum choice can be made by minimization of the mean value of
$<0|H|0>$,  but it is no so for time-dependent oscillator.

 However let us remind that when $\omega=conat$
the mean value of an observable oscillates with time for a
squeezed vacuum state, whereas it is zero or constant for a true
vacuum state. One can suppose, that this is the clue to issue of
the true vacuum state of time-dependent oscillator. Namely, an
observable value oscillates with time for a squeezed vacuum state,
but it is monotonic function for a true vacuum state. It is
convenient to choose the difference of the oscillator kinetic and
potential energies to be this observable. For the vacuum state of
oscillator with constant frequency, this quantity equals to zero
according to the virial theorem and we will see that this quantity
is a monotonic function of time for a vacuum state of
time-dependent oscillator (if such a state exists).

\section{Time-dependent oscillator: examples of the vacuum states}

We intend to concentrate the different examples, which grade with
the asymptotic of the adiabatic parameter ${\dot
\omega}/{\omega^2}$: i) $\omega \rightarrow const$, ${\dot
\omega}/{\omega^2}\rightarrow 0$; ii) $\omega$ does not tends to a
constant but ${\dot \omega}/{\omega^2}\rightarrow 0$; iii) then
${\dot \omega}/{\omega^2}=const$, and, at last, iv) ${\dot
\omega}/{\omega^2}\rightarrow \infty$.

 Let $\omega$ depends on $t$ in the well-known
form \cite{birrel}:
\begin{equation}
\omega(t)=k\sqrt{1+\tanh(H \,t).} \label{dep1}
\end{equation}

The solution of Eqs. (\ref{rel0}) and (\ref{osc}) is
\cite{birrel}:
\begin{eqnarray}
u_0=2^{-3/4}
   {k^{-1/2}}\mbox{e}^{-\frac{i k t}{\sqrt{2}}} \left(\mbox{e}^{-H
t}+\mbox{e}^{H t}\right)^{-\frac{i k}{\sqrt{2} H}}~~~~~~~~~ \nonumber\\
_2F_1\left(\frac{\mbox{i} k}{\sqrt{2}
   H},\frac{\mbox{i} k}{\sqrt{2} H}+1;\frac{\mbox{i}
    \sqrt{2} k}{H}+1;\frac{1}{1+\mbox{e}^{2 H t}}\right)
, \label{u0}
\end{eqnarray}
where $_2F_1(a,b;c;z)$ is the  hypergeometric function \cite{ab}.
An arbitrary solution from a family of the squeezed vacuum states
is given by Eqs. (\ref{bog}).

 Mean value of the kinetic and potential energies difference
is expressed as
\begin{equation}
<0|\frac{1}{2}p^2-\frac{1}{2}\omega x^2|0>=\frac{1}{2}({\dot u}
{\dot u}^{*}- \omega^2 u u^* )=\dot \sigma(t).
\end{equation}
Here
\begin{equation}
\sigma=\frac{1}{2}<0|\hat x\hat p+\hat p\hat x|0>=\frac{1}{2}(\dot
u u^*+{\dot u}^{*}u)
\end{equation} has
a sense of additional uncertainty arising in the Heisenberg
uncertainty relation \cite{hez}:
\begin{equation}
<|(\hat p-p_0)^2|><|(\hat x-x_0)^2|>~ > 1/4+\sigma^2,
\label{inecv}
\end{equation}
where $p_0=<\hat p>$, $x_0=<\hat x>$ and $|>$ is the arbitrary
state. For a family of the squeezed vacuum states, including the
true vacuum, the inequality (\ref{inecv}) becomes an equality.

\begin{figure}[h]
\hspace {-1.5 cm} \epsfxsize =7. cm \epsfbox{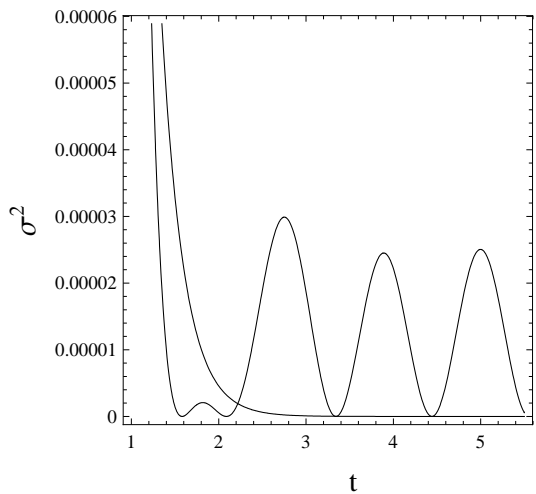}
\caption{The function $\sigma^2$ for the dependence of $\omega(t)$
given by (\ref{dep1}) and $k=1$, $H=1$. Monotonic curve
corresponds to the vacuum state ($r=0$, $\delta=0$ ) and
oscillating curve corresponds to the squeezed state ($r=0.005$,
$\delta=0$).}
 \vspace{0.0cm}
\label{s}
\end{figure}

Fig. \ref{s} shows the $\sigma(t)$-function for different values
of the parameters $r,\delta$. One can see that this function
oscillates at some value of the parameters $r$, $\delta$ and the
only parameter $r=0$ results in the monotonic behavior of
$\sigma(t)$.

It should be noted that the selection rule have been offered
\cite{kim} for a vacuum state as a state having the minimal
uncertainty at each moment of time. As one can see from the above
example, this rule is not satisfied for the vacuum state. Indeed,
there exists a region in Fig. {\ref{s}}, where the uncertainty for
the slightly squeezed state is less than that for the vacuum
state. Hence, one has to conclude that this selection rule is not
valid in the general case.

\begin{figure*}[t]
\hspace {-0.5 cm} \epsfxsize =13. cm \epsfbox{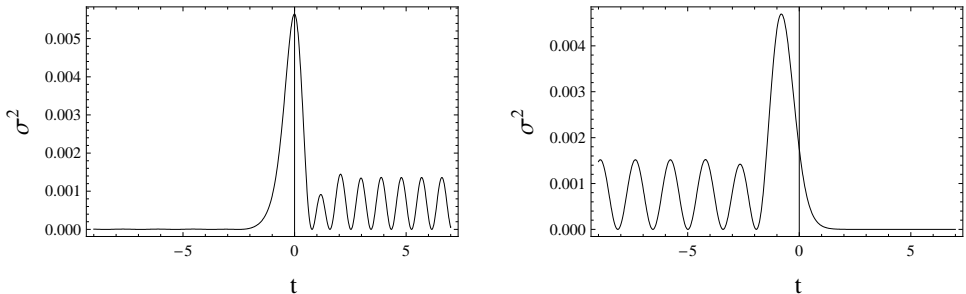}
\caption{The function $\sigma^2$ for -in and -out vacuum states.
The dependence $\omega(t)$ is given by (\ref{dep11}) and $k=1$,
$H=1$.}
 \vspace{-0.5cm}
\label{sa}
\end{figure*}

Our suggestion is to correlate a vacuum state with the monotonic
time-dependence of the functions $\sigma$ or $\dot \sigma$. That
is, for this example, the true vacuum corresponds to the function
$u_0$ given by (\ref{u0}). As the criterium for choosing the
function with monotonic behavior, one can use the minimization of
the functional
\begin{equation}
Z(r,\delta)=\lim_{T\rightarrow \infty}\left(\int_{t_0}^T
(\partial_t \sigma(t,r,\delta))^2 \mbox{d} t \over{\int_{t_0}^T
(\partial_t \sigma(t,r_0,\delta_0))^2 \mbox{d} t}\right),
\label{func}
\end{equation}
where $r_0, \delta_0$ are some fixed values used for
normalization.
  The exact analytic calculation of the functional  with the function
$u$ from (\ref{bog}), (\ref{u0}) and values $r_0=\ln 2$,
$\delta_0=0$ gives
\begin{equation}
Z=\frac{64}{225}\sinh^2 r. \label{z}
\end{equation}
Thus, the minimization of the functional leads to the value $r=0$
for the vacuum state. This is because the function $\sigma$ has
the asymptotic
\begin{equation}
\sigma(t)\approx-\frac{1}{2} \sin \left(2 \sqrt{2} k \,t+\delta
\right) \sinh (2 r)
\end{equation}
at infinity.

Instead of the parameters $r$ and $\delta$, one can seek the
initial conditions for Eq. (\ref{osc}) at some $t_0$. Really, the
represention $u(t)=\mbox{e}^{\mbox{i}\varphi(t)}\theta(t)$ leads
to $\dot \varphi=-\theta^{-2}/2$ from Eq. (\ref{rel0}). That is,
$\theta(t_0)$ and $\dot \theta(t_0)$ define $u(t_0)$ and $\dot
u(t_0)$ completely because the phase $\varphi(t_0)$ can be chosen
to be zero. Then one can solve Eq. (\ref{osc}) with some initial
conditions and find the value of the functional (\ref{func}).
Initial conditions giving the minimum of the functional correspond
to the vacuum state.

Moreover, one can write the differential equation directly for
$\sigma(t)$ \cite{conf}.

Straightforward computation shows that $\sigma$ satisfies the
equation
\begin{equation}
\dddot\sigma -\ddot\sigma  \left(\frac{\dot \omega }{\omega
}+\frac{\ddot\omega }{\dot\omega }\right)+4 \dot\sigma \omega
^2+\sigma
   \left(8 \omega  \dot\omega -\frac{4 \omega ^2\ddot \omega }{\dot\omega}\right)=0, \label{diffsig}
\end{equation}
if $u(t)$ obeys (\ref{osc}).

  The relation (\ref{rel0}) leads to
\begin{equation}
\frac{1}{\omega\dot\omega^2}\left(4 \sigma \omega
^2+{\ddot\sigma}\right) \left(4 \sigma \,\omega ^3+{\ddot\sigma}\,
\omega -2 {\dot\sigma}\dot \omega \right)-4 \sigma^2 =1,
\label{unssig}
\end{equation}
for the states belonging  to a family of the squeezed vacuum
states including the true vacuum.
    Left hand side of Eq. (\ref{unssig}) is the integral of motion of (\ref{diffsig}).

 One can connect the initial condition for Eq. (\ref{diffsig}) with that for Eq. (\ref{osc}):
\begin{eqnarray}
\sigma=\theta \,
 \dot\theta,~~~~~~~~~~~~~~~~~~\\ \dot\sigma={\dot\theta}^{
2}+\theta^2({\dot \varphi}^{2}-\omega^2).
\end{eqnarray}

Second derivative of $\sigma$ can be expressed through Eq.
(\ref{unssig}). Thus, the determination of $\sigma(t_0)$ and
$\dot\sigma(t_0)$ allows solving Eq. (\ref{diffsig}) instead of
defining $\theta(t_0)$ and $\dot\theta(t_0)$ and solving Eq.
(\ref{osc}).

In the above example, the function $\sigma$  has the monotonic
behavior in the vacuum state within all range of $t$. This means
that the single global vacuum exists. The more complicated case
\cite{birrel} is
\begin{equation}
\omega(t)=k\sqrt{2+\tanh(H \,t)} \label{dep11},
\end{equation}
where there are two different non zero values of  $\omega$
 at $t\rightarrow \infty$ and $t\rightarrow -\infty$.

One can see from Fig. \ref{sa}, that two vacuum solutions exist.
One of them has the monotonic behavior at $t\rightarrow +\infty$
(out-vacuum state) and the second one has such a behavior at
$t\rightarrow -\infty$ (in-vacuum state ). In this paper we do not
discuss an important issue concerning the dependencies $\omega(t)$
providing the unique global vacuum \cite{footnote1}, but if in- or
out- vacuums exist, the out-vacuum can be found by the
minimization of the functional (\ref{func}), whereas the in-vacuum
state can be found by setting $T\rightarrow - \infty$ in
(\ref{func}).

The considered cases are simple in the sense that $\omega$ tends
to a constant and a notion of particle is asymptotically defined.
For example, the out-vacuum means an absence of the particles at
$t\rightarrow \infty$ and, simultaneously, the function $\sigma$
has the monotonic behavior.

\begin{figure}[h]
\hspace {0.0 cm} \epsfxsize =7. cm \epsfbox{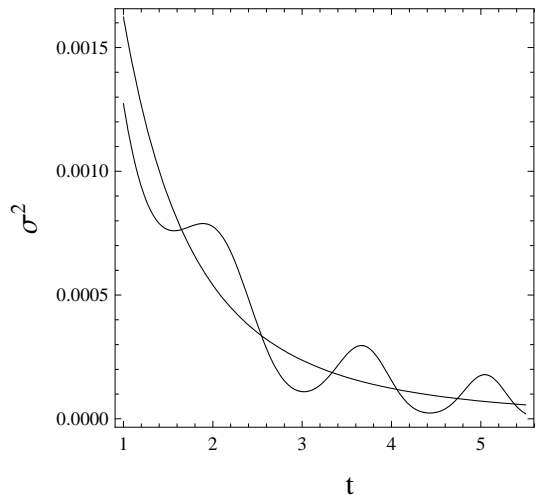} \caption{The
function $\sigma^2$ for the dependence $\omega(t)$ given by
(\ref{dep22}) and $k=1$, $H=1$. Monotonic curve corresponds to the
vacuum state ($r=0$, $\delta=0$ ) and oscillating curve
corresponds to the squeezed state ($r=0.005$, $\delta=0$).}
 \vspace{-0.5cm}
\label{s11}
\end{figure}

Now let us consider the example
\begin{equation}
\omega(t)=\sqrt{1+H t}, \label{dep22}
\end{equation}
where $\omega(t)$ does not tend to a constant at infinity,
 but the adiabatic condition  $\left|\frac{\dot\omega}{\omega^2}\right|\rightarrow 0 $ is still satisfied
 at $t\rightarrow\infty$. Eqs. (\ref{osc}), (\ref{rel0}) are solvable in the closed   form
\begin{equation}
u_0={H^{-1/6}k^{-1/3}}{\sqrt{2 \pi }\,
\mbox{Ai}\left(\frac{2k^{2/3}(H
t+1)}{H^{2/3}\left(1-i\sqrt{3}\right) }\right)},
\end{equation}
where $\mbox{Ai}(z)$ denotes the Airi function \cite{ab}. Again,
the calculation of the functional (\ref{func}) gives Eq.
(\ref{z}). It should be noted that the asymptotic of $\sigma$ is
(see also Fig. \ref{s11})
\begin{eqnarray}
\sigma\approx-\frac{1}{4} \biggl(\cos \left(\frac{4 k (H
t+1)^{3/2}}{3 H}+\delta \right)~~~~~~~~\nonumber\\+\sqrt{3} \sin
\left(\frac{4 k (H t+1)^{3/2}}{3
   H}+\delta \right)\biggr) \sinh (2 r).
\end{eqnarray}

Let us come to the example, where the adiabatic condition is not
fulfilled:
\begin{equation}
\omega(t)=\frac{k}{1+2 H t}. \label{onemore}
\end{equation}
The adiabatic parameter
$\bigl|\frac{\dot\omega}{\omega^2}\bigr|=\frac{2 H}{k}$ is
constant. However, as it will be shown, the vacuum state exists in
this case, as well. The solution of Eqs. (\ref{osc}), (\ref{rel0})
is
\begin{equation}
u_0=\frac{(1+2 H t)^{-{\mbox{i}} \sqrt{k^2-H^2}/(2 H
)+{1}/{2}}}{\sqrt{2} (k^2-H^2)^{1/4}},
\end{equation}
and results in the following  asymptotic
\begin{figure}[t]
\hspace {0.0 cm} \epsfxsize =7. cm \epsfbox{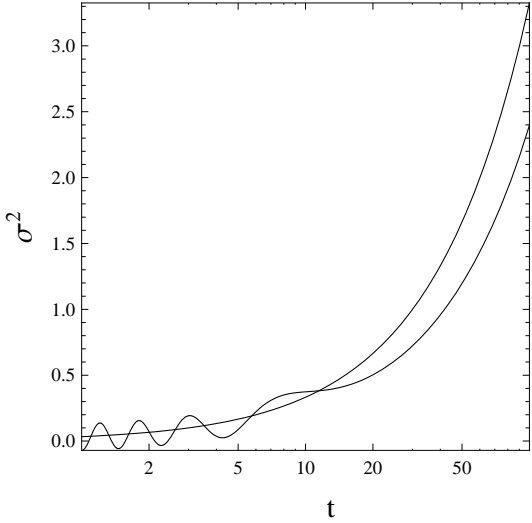} \caption{The
function $\sigma^2$ for the dependence $\omega(t)$ given by
(\ref{dep33}) and $k=1$, $H=1$. Monotonic curve corresponds to the
vacuum state ($r=0$, $\delta=0$ ) and oscillating curve
corresponds to the squeezed state ($r=0.005$, $\delta=0$).}
 \vspace{0. cm}
\label{s12}
\end{figure}

\begin{widetext}
\begin{eqnarray}
\sigma\approx\frac{1}{2\sqrt{k^2-H^2}}\Biggl({H \cosh ^2(r)}+{H
\sinh
   ^2(r)}-\sqrt{k^2-H^2}\sin \biggl(\delta+\frac{1}{H}{\sqrt{k^2-H^2} \ln (2 H t+1)}\biggr)
    \sinh (2r)\nonumber \\+{H  \cos \left(\delta+\frac{1}{H}{\sqrt{k^2-H^2}
   \ln (2 H t+1)}\right) \sinh (2 r)}\Biggr).
\end{eqnarray}
\end{widetext}

The comparison with the previous cases demonstrates that the
constant component appears in the asymptotic \cite{footnote2}.
However, again the functional (\ref{func}) has the form (\ref{z})
for $k>H$. In the opposite case the function $\sigma$ has
non-oscillating behavior at infinity under arbitrary initial
conditions.

Now let us consider the following example:
\begin{equation}
\omega=\frac{ k}{1+ H^2 t^2}, \label{dep33}
\end{equation}
where the adiabatic parameter
$\left|\frac{\dot\omega}{\omega^2}\right|=\frac{H^2 t}{k}$ becomes
greater than unity at large $t$.
 The expression for the function $u_0$ has the form

\begin{equation}
u_0=\frac{\sqrt{H^2 t^2+1}}{\sqrt{2}
\sqrt[4]{H^2+k^2}}\exp=\left({-\frac{\mbox{i}}{H}{
\sqrt{H^2+k^2}\, \arctan(H t)}} \right),
\end{equation}
as well as the expression for $\sigma$ is

\begin{widetext}
\begin{equation}
 \sigma=\frac{t \cosh (2 r) H^2}{{2
\sqrt{H^2+k^2}}}+\Biggl(\frac{H^2 t}{{2 \sqrt{H^2+k^2}}} \cos
\biggl(\delta +\frac{2 \sqrt{H^2+k^2} \arctan(H
t)}{H}\biggr)-\frac{1}{2}\sin
   \biggl(\delta +\frac{2 \sqrt{H^2+k^2} \tan ^{-1}(H t)}{H}\biggr)\Biggr) \sinh (2
   r).
\end{equation}
\end{widetext}

In this example, the function $\sigma(t)$ for an arbitrary state
has non-oscillating behavior at infinity (Fig. \ref{s12}). It
occurs because the function $u$ itself ceases to oscillate at
$t\rightarrow \infty$. In the literature \cite{lin}, such a
phenomenon is interpreted as the transition from the quantum field
to the classical one. The behavior of $\sigma$ confirms this
interpretation because the absence of oscillations means the
absence of interference (i.e., in fact, absence of the main
constituent of quantum mechanics). In any case we cannot talk
about an existence of the out- vacuum state here. However, one can
introduce a concept of the approximate vacuum state corresponding
to some range of $t$. One can see from Fig. \ref{s12}, that there
exists a range, where the typical non-vacuum $\sigma$ oscillates
and thus, an approximate vacuum state corresponding to the
non-oscillating $\sigma$ can be defined.

\section{Vacuums of the scalar field oscillator}

In principle a number of the approximate vacuums corresponding to
the
 different time regions can exist. Let us take an example, which does not
admit an analytical consideration.

Lagrangian corresponding to the  modes of the scalar field in an
expanding Universe has the form \cite{birrel}
\begin{equation}
\mathcal{ L}_{scal}=\frac{1}{2}\sum_{\boldsymbol k}
a^2{\phi^\prime_{\boldsymbol k}\phi^\prime_{-{\boldsymbol k}}}-
a^2 k^2 \phi_{\boldsymbol k}\phi_{-{\boldsymbol k}} - a^4 m^2
\phi_{\boldsymbol k}\phi_{-{\boldsymbol k}},
\end{equation}
where $\phi_{\boldsymbol k}$ is the  Fourier-transform of the
scalar field $\phi(\boldsymbol r)=\sum_{\boldsymbol k}
\phi_{\boldsymbol k} \mbox{e}^{i {\boldsymbol k}\boldsymbol r}$
and $a(\tau)$ is the scale factor of Universe, $\tau$ is the
conformal time $dt=a(\tau)d\tau $.

The equation of motion can be deduced

\begin{equation}
\phi^{\prime\prime}_{\boldsymbol k}+(k^2+a^2 m^2)\phi_{\boldsymbol
k}+2\frac{ a^\prime}{a}{  \phi^\prime}_{\boldsymbol k}=0 .
\end{equation}

Quantization of the scalar field \cite{birrel}
\begin{equation}
\hat \phi_{\boldsymbol k}=\hat {\mbox{a}}^+_{-\boldsymbol
k}\chi_{k}^*(\tau)+\hat {\mbox{a}}_{\boldsymbol k} \chi_{k}(\tau)
\end{equation}
leads to the operators of creation and annihilation with the
commutation rules
 $[{\hat{\mbox{a}}}_{\boldsymbol k},\, {\hat{\mbox{a}}}^+_{\boldsymbol k}]=1$. The complex functions
 $\chi_k(\tau)$
satisfy the relations \cite{birrel}:
\begin{eqnarray}
\chi^{\prime\prime}_k+(k^2+m^2 a^2) \chi_k+2\frac{ a^\prime}{a}{
\chi^\prime}_k=0,\nonumber\\
a^2(\tau)(\chi_k
\,{\chi_k^\prime}^*-\chi_k^*\,\chi_k^\prime)=\mbox{i}. \label{rel}
\end{eqnarray}

Substitution of $\chi_{k}=u_{k}/a$ results in the time dependent
oscillator:
\begin{equation}
u_k^{\prime\prime}+(k^2+a^2
m^2+\frac{a^{\prime\prime}}{a})u_{k}=0.
\end{equation}

\begin{figure}[h]
\hspace {0.0 cm} \epsfxsize =7. cm \epsfbox{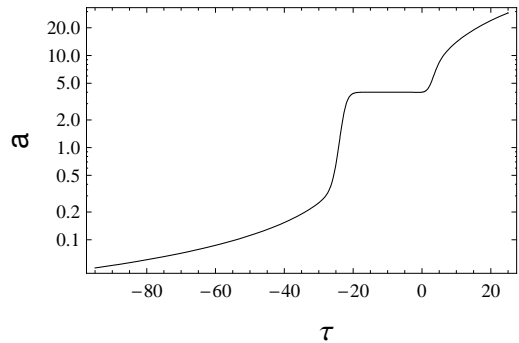} \caption{The
dependence (\ref{atau}) of the Universe scale factor $a(\tau)$ on
the conformal time $\tau$.}
 \vspace{0.5cm}
\label{s14}
\end{figure}

\begin{figure*}[t]
\hspace {0.0 cm} \epsfxsize =15. cm \epsfbox{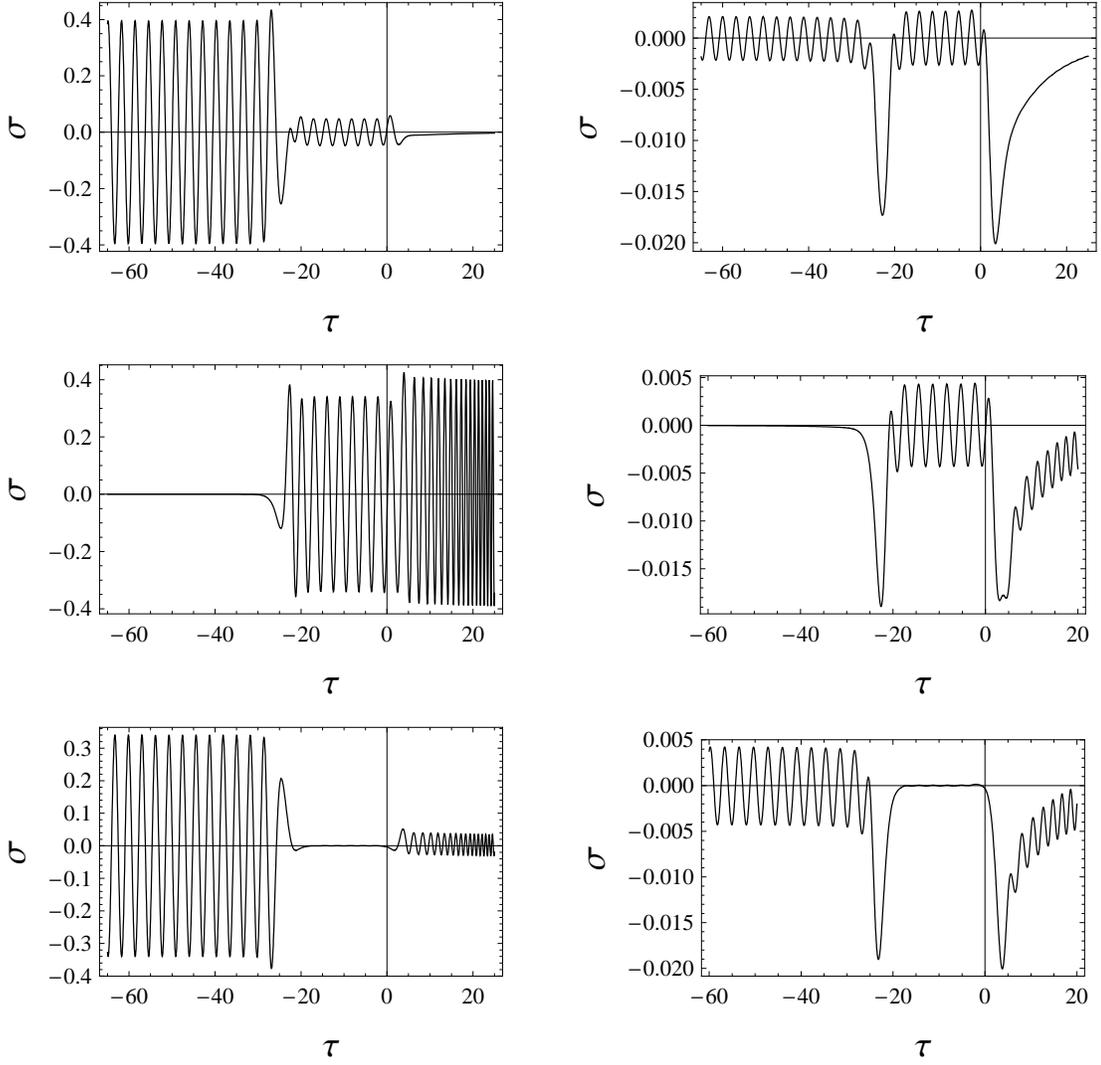} \caption{
The function $\sigma$ for the vacuum states inspired by the
dependence (\ref{atau}) and $k=1$, $m=1/16$. Left column
corresponds to the scalar field oscillator, whereas right column
corresponds to the fermionic one. Upper, middle and bottom rows
correspond to the solutions representing the vacuums for the out-,
in-, and central-time-ranges, respectively. }
 \vspace{0. cm}
\label{s15}
\end{figure*}

Now we consider some illustrative time-dependence $a(\tau)$
\begin{equation}
a(\tau)=\tau
\left(1+\exp(3-\tau)\right)^{-1}+4\left(1-\frac{\tau+15}{1+\exp(\tau+25)}\right)^{-1},
\label{atau}
\end{equation}
which is shown in Fig. \ref{s14}. There exist three ranges, where
one can search for the vacuum state. Namely, one can try to find
the in-, out- vacuum states and, besides, the approximate vacuum
state for the central range (from $\approx -$20 to $0$ shown in
Fig. \ref{s14}). The numerical minimization of the functional
\begin{equation}
Z(\alpha,\beta)=\int_{\tau_1}^{\tau_2} (\partial_\tau
\sigma(\tau,\alpha,\beta))^2 \mbox{d}\tau, \label{func1}
\end{equation}
where $\sigma(\tau)$ obeys (\ref{diffsig}), (\ref{unssig}) and
$\alpha=\sigma(\tau_1)$, $\beta=\sigma^\prime(\tau_1)$, allows
finding the initial conditions corresponding to the vacuum. The
solutions are shown in Fig. \ref{s15}.

\section{Vacuums of the fermionic oscillator}

Let us come to the fermionic oscillator. After decomposition of
the bispinor $\psi(\boldsymbol r)$ in the complete set of modes
 $\psi(\boldsymbol
r)=\sum_{\boldsymbol k} \psi_{\boldsymbol k} \mbox{e}^{\mbox{i}
{\boldsymbol k}\boldsymbol r}$, Lagrangian of the fermion field in
the expanding Universe (see \cite{el, kof,pel,2} and reference
therein) takes the form
\begin{eqnarray}
 L=\sum_{\boldsymbol k}\frac{\mbox{i}\, a^3 }{2}\psi^+_{\boldsymbol
k}\partial_\tau\psi_{\boldsymbol k}-\frac{\mbox{i}\, a^3}{2}
\partial_\tau \psi^+_{\boldsymbol k}\psi_{\boldsymbol k}-a^3\psi^+_{\boldsymbol k}
(\boldsymbol \alpha \boldsymbol k)\,\psi_{\boldsymbol
k}\nonumber\\-a^4\, m \psi^+_{\boldsymbol k}\beta
\psi_{\boldsymbol k}.
\end{eqnarray}

The equation of motion is
\begin{equation}
i{\psi}^\prime_k- (\boldsymbol \alpha  \boldsymbol
k){\psi}_{\boldsymbol k} +\mbox{i}
\frac{3a^\prime}{2a}{\psi}_{\boldsymbol k}-m\,a
\beta{\psi}_{\boldsymbol k}=0, \label{rr}
\end{equation}
Fermion field is quantized as
\begin{equation}
\hat \psi_{\boldsymbol k}={\hat b}^+_{-\boldsymbol k,s}{
v}_{-\boldsymbol k,s}+{\hat a}_{\boldsymbol k,s}u_{\boldsymbol
k,s},
\end{equation}
where the bispinor is \cite{footnote3}:
\[u_{\boldsymbol k,s}(\eta) = \frac{\mbox{i}\chi_k^\prime+m a\chi_k}{a^{3/2}}\left (
\begin {array}{c}
 \varphi_s \\
\frac{\chi_k(\boldsymbol \sigma \boldsymbol k)}{i\chi_k^\prime+m
\chi_k a}\varphi_s
\end {array}
\right)~,
\]
and spinors $\varphi_s$ are
 $\varphi_+=\left (
\begin {array}{c}
1 \\
0
\end {array}
\right)$ and $\varphi_-=\left (
\begin {array}{c}
0 \\
1
\end {array}
\right)$.

The bispinor $v_{\boldsymbol k,s}$  is expressed as
$v_{\boldsymbol k,s}=\mbox{i}\gamma^0\gamma^2(\bar u_{\boldsymbol
k,s})^{T}$, where the symbol $T$ denotes the transpose vector and
$\bar u=u^+\gamma^0$. The functions $\chi_{k}(\eta)$ satisfy
\cite{2}

\begin{eqnarray}
\chi^{\prime\prime}_{k }+ \left(k^2+m^2 a^2-\mbox{i} m
a'\right)\chi_{ k}=0, \label{1}
\\
 k^2\chi_{ k}  \chi^*_{ k} +\left(a m \chi^*_{k}-\mbox{i} {\chi
^\prime_{k}}^*\right) \left(a m \chi_{k} +\mbox{i} \chi^\prime_{
k} \right)=1, \label{rel1}
\end{eqnarray}
and again there appears time-dependent oscillator (with the
complex frequency), where the functions $\chi_k$ plays a role of
the above mentioned $u_k$. The true vacuum state can be defined as
that providing a non-oscillating behavior of the function
\begin{equation}
\sigma_k(t)=\frac{1}{2}\left(\chi^*_k\chi_k^\prime+\chi_k^{\prime
*}\chi_k\right).
\end{equation}
One can deduce that if $\chi_k(t)$ obeys (\ref{1}) then $\sigma_k$
satisfy
\begin{eqnarray}
 \sigma_k^{\prime\prime\prime}-\frac{{\sigma_k^{\prime\prime}}
M''}{M'}+4(k^2+M^2){\sigma_k^\prime}~~~~~~~~~~~~~~~~~~~~~~~~~~~\nonumber
\\+\left(12 { M
   M'}-\frac{4 M''
k^2}{M'} -\frac{4  M^2 M''}{M'}\right){\sigma_k }=0,
\end{eqnarray}
where $M(\tau)=m a(\tau)$. The relation (\ref{rel1}) gives
\begin{eqnarray}
\frac{1}{M^{\prime 2}}\left(k^2+M^2\right) \left(4 \sigma_k k^2+4
\sigma_k M^2+\sigma_k ''\right)^2~~~~~~~~~~~~~~~\\ \nonumber-2
\frac{M}{M'} \sigma_k '
 \left(4 \sigma_k k^2+4
   \sigma_k  M^2+\sigma_k ''\right)+4 k^2 \sigma_k ^2+\sigma_k^{\prime 2}=1.
\end{eqnarray}
 The vacuum solutions obtained by minimization of the functional (\ref{func1})
 in the three different ranges are shown in
 Fig. \ref{s15} (right column).

\section{Vacuums of the two coupled oscillators with constraint}

Now we address ourself to a little more complicated system:
namely, the two time-dependent coupled oscillators with
constraint. This system appears in the theory of anisotropy of the
Cosmic Microwave Background \cite{muh,sas}. One can expect, that
some difficulties will arise with the vacuum definition for this
system, because the quantization of constrained systems can
reveals some nontrivial features. However, we will see that there
are no pathologies in this particular case.

Both scalar field and gravitation  can be assumed to be specified
by the action \cite{birrel,lin}
\begin{equation}
S=-\frac{1}{16\pi G}\int \mbox{d}^4 x\sqrt{-g}R+\int \mbox{d}^4
x\sqrt-g[ (\frac{1}{2}\partial_\mu\phi)^2-V(\phi)], \label{deistv}
\end{equation}
Representation of the metric tensor in the form \cite{muh,lin}
\begin{equation}
\mbox{d} s^2=(1+2 \Phi(\boldsymbol r,t))\mbox{d} t^2-a^2(t)(1-2
\Phi(\boldsymbol r,t))\mbox{d}{\boldsymbol r}^2,
\end{equation}
and considering the scalar field as that possessing a spatially
uniform component with a small perturbation around it:
\begin{equation}
\phi(\boldsymbol r, t)=\phi(t)+\theta(\boldsymbol r,t)
\end{equation}
allows obtaining the system of equations \cite{lin} of zero order
in $\theta$ and $\Phi$,

\begin{eqnarray}
-{{\dot a}^2}{a}+{\dot\phi}^2a^3+2 a^3\,V(\phi)=0,\label{zeroa}\\
\ddot a=-\frac{3}{2}a {\dot\phi}^2-\frac{{\dot a}^2}{2 a}+{3} a V(\phi),\label{zerob}\\
\ddot \phi+3\frac{\dot a}{a}\dot \phi+\frac{\mbox{d} V
}{\mbox{d}\phi}=0,\label{zeroc}
\end{eqnarray}

where we use system of units ${4\pi G}/{3}=1$. The first order
equations for the Fourier-transformed perturbations of the scalar
field $\theta(\boldsymbol r,t)=\sum_{\boldsymbol k}
\theta_{\boldsymbol k}(t) \mbox{e}^{\mbox{i} {\boldsymbol
k}\boldsymbol r}$ and metric $\Phi(\boldsymbol
r,t)=\sum_{\boldsymbol k} \Phi_{\boldsymbol k}(t)
\mbox{e}^{\mbox{i} {\boldsymbol k}\boldsymbol r}$ have the
following form \cite{lin,muh}:
\begin{eqnarray}
\frac{1}{3}\Phi_{\boldsymbol k}  k^2+ \frac{\mbox{d} V}{\mbox{d}
\phi} \theta_{\boldsymbol k} a^2 +{\dot\Phi_{\boldsymbol k} \dot
a}{a}+\dot \theta_{\boldsymbol k} \dot\phi a^2+2 \Phi_{\boldsymbol
k} a^2V(\phi)=0,\nonumber\\ \label{c1}
\\
-\frac{1}{3} \dot \Phi_{\boldsymbol k}-\frac{\Phi_{\boldsymbol k} \dot a}{3 a}+\theta_{\boldsymbol k} \dot \phi=0,~~~\label{c2}\\
-3 \frac{\mbox{d} V}{\mbox{d} \phi}\theta_{\boldsymbol k} -\ddot
\Phi_{\boldsymbol k}-4\dot \Phi_{\boldsymbol k}\frac{\dot
a}{a}+3\dot
\theta_{\boldsymbol k} \dot \phi-6 V( \phi) \Phi_{\boldsymbol k}=0,~~~ \label{c3}\\
\ddot \theta_{\boldsymbol k}+3\frac{\dot a}{a}\dot
\theta_{\boldsymbol k}+\frac{k^2}{a^2}\theta_{\boldsymbol
k}+\frac{\mbox{d}^2 V}{\mbox{d}\phi^2}\theta_{\boldsymbol k}+2
\frac{\mbox{d} V }{\mbox{d}
\phi}\Phi_{\boldsymbol k}-4 \dot \phi \dot \Phi_{\boldsymbol k}=0. \nonumber \\
\label{eqns}
\end{eqnarray}

Eqs. (\ref{c3}), (\ref{eqns}) are the equations of motion. They
can also be obtained from Lagrangian

\begin{widetext}
\begin{eqnarray}
L=\sum_{\boldsymbol k}-\frac{1}{2}\frac{\mbox{d}^2 V}{\mbox{d}
\phi^2}\theta_{\boldsymbol k}\theta_{-\boldsymbol k}a^3+2
\frac{\mbox{d} V }{\mbox{d} \phi }\theta_{\boldsymbol k}
\Phi_{-\boldsymbol k} a^3-10 V \Phi_{\boldsymbol
k}\Phi_{-\boldsymbol k}a^3-\frac{1}{2} a k^2\theta_{\boldsymbol
k}\theta_{-\boldsymbol k}~~~~~~~~~~~~~~~~~~~~~~~~~~~
\nonumber\\
-\frac{1}{6}ak^2\Phi_{\boldsymbol k}\Phi_{-\boldsymbol
k}+\frac{1}{2}a^3{\dot\theta_{\boldsymbol
k}}{\dot\theta_{-\boldsymbol k}} -\frac{1}{2}a^3{\dot
\Phi_{\boldsymbol k}}{\dot \Phi_{-\boldsymbol k}} -4
\Phi_{\boldsymbol k} \dot \Phi_{-\boldsymbol k} a^2\dot a-4
a^3\Phi_{\boldsymbol k}  {\dot \theta_{-\boldsymbol k}}\dot \phi.
\end{eqnarray}
\end{widetext}

Eqs. (\ref{c1}), (\ref{c2}) are the constraints. However, Eq.
(\ref{c1}) can be derived from Eqs. (\ref{c2}), (\ref{c3}),
(\ref{eqns}) and, thus, it is not independent. That is there are
two time dependent oscillators with one constraint of the first
kind \cite{git}. Using this constraint one can exclude the scalar
field perturbation from (\ref{c3}) and obtain
\begin{eqnarray}
 \ddot \Phi_{\boldsymbol k}-\dot \Phi_{\boldsymbol k} \frac{\mbox{d}}{\mbox{d}
t}\ln\left(\frac{{\dot a}^2}{a^3}-\frac{\ddot
a}{a^2}\right)+\biggl(\frac{k^2}{a^2}+\frac{2 \ddot a }{a}-\frac{2
{\dot a}^2}{a^2}~~~~~~~\nonumber\\-\frac{\dot
a}{a}\frac{\mbox{d}}{\mbox{d} t}\ln\left( \frac{{\dot
a}^2}{a^2}-\frac{\ddot a}{a}\right)\biggr)\Phi_{\boldsymbol k}=0,
\label{eqphi}
\end{eqnarray}
where the uniform scalar field $\phi$ has been excluded by using
(\ref{zeroa}), (\ref{zerob}) as well. Quantization consists in

\begin{equation}
\hat \Phi_{\boldsymbol k}=\hat {\mbox{a}}^+_{-\boldsymbol
k}u_{k}^*(t)+\hat {\mbox{a}}_{\boldsymbol k} u_{k}(t),
\end{equation}
and the vacuum can be found by minimization of the quantity
$Z(\alpha_k,\beta_k)$ (\ref{func1}), that allows finding
$\alpha_k$ and $\beta_k$, which correspond to the vacuum state. In
this we have solved Eq. (\ref{eqphi}) directly and have written
the initial conditions at $t_0$  as $u_k(t_0)=\alpha_k $, $\dot
u_k(t_0)=\beta_k-\mbox{i}\frac{ X(t_0)}{2 \alpha_k }$, where
$X(t_0)=\frac{{\dot a}^2}{a^3}-\frac{\ddot a}{a^2}\bigr|_{t=t_0}$.
These initial conditions are consistent with the relation
\begin{equation}
u_k {\dot u_k}^*-u_k^*\dot u_k=\mbox{i} X(t),
\end{equation}
which is analog of (\ref{rel}) and corresponds to the general case
of quantization of oscillator with the time-dependent mass and
frequency \cite{kim1}.

\begin{figure}[h]
\hspace {0.0cm} \epsfxsize =8. cm \epsfbox{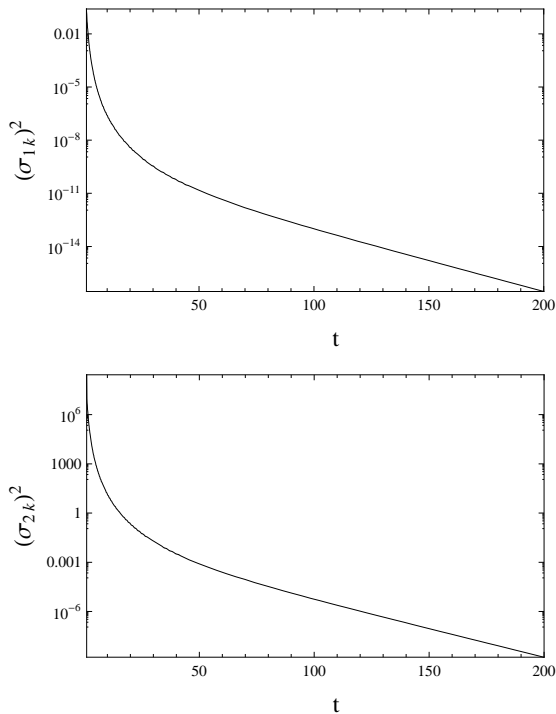} \caption{The
function $\sigma^2$ corresponding to the vacuum for
$a(t)=\sinh(\gamma t)$, where $\gamma=\frac{1}{50}$  and $k=1$.
top panel shows result of the metric perturbation quantization,
bottom panel shows the result of the scalar field perturbation
quantization.}
 \vspace{-0.5cm}
\label{s16}
\end{figure}

On the other hand, one can express $\Phi_{\boldsymbol k}$ through
$\theta_{\boldsymbol k}$, and obtain the equation for
$\theta_{\boldsymbol k}$ analogously to (\ref{eqphi}). This
equation turns out to be more complicated and we do not write it
here. The question arises: would be the vacuum state the same, if
the scalar field perturbation is quantized as:
\begin{equation}
\hat \theta_{\boldsymbol k}=\hat {\mbox{a}}^+_{-\boldsymbol
k}{\mathcal U}_{k}^*(t)+\hat {\mbox{a}}_{\boldsymbol k} {\mathcal
U}_{k}(t)\, ?
\end{equation}

From Eq. (\ref{c2}), we have
\begin{equation}
{\mathcal U}_k =\frac{1}{3\dot  \phi}\left(\dot u_k+\frac{\dot
a}{a}u_k \right), \label{U}
\end{equation}
where Eqs. (\ref{zeroa}), (\ref{zerob}) reduce $\dot \phi$ to the
form $\dot \phi =\sqrt{\frac{1}{3}\frac{{\dot
a}^2}{a^2}-\frac{1}{3}\frac{\ddot a }{a}}$.

Let us consider the particular case of $a(t)=\sinh(\gamma t)$. The
numerical minimization of the functional allows finding the vacuum
solution for which the function $\sigma_{1k}=\frac{1}{2}(\dot u_k
u_k^*+{\dot u_k}^{*}u_k)$ has monotonic behavior and, thus,
corresponds to the vacuum state. Now if one expresses
$\sigma_{2k}=\frac{1}{2}(\dot {\mathcal U}_k {\mathcal
U}_k^*+{\dot {\mathcal U}_k}^{*}{\mathcal U}_k)$ through the
functions ${\mathcal U}_{\boldsymbol k}$ given by (\ref{U}), it is
seen from Fig.\ref{s16}, that $(\sigma_{2k})^2$ has also monotonic
behavior. Thus, it is the vacuum state for the
$\theta_{\boldsymbol k}$-oscillator too. Moreover, one can choose
any convenient variable from a combination of $\Phi_{\boldsymbol
k}$ and $\theta_{\boldsymbol k}$ as it usually done
\cite{muh,sas,bard}.

Procedure "NMinimize" of the Wolfram software "Mathematica" is
used in all the numerical calculations.

\section{Conclusion}

We have considered the method to find the vacuum state of a driven
quantum oscillator numerically by the means of minimization of the
functional containing the square of derivative of the additional
uncertainty $\sigma$ arising in the Heisenberg uncertainty
relation. For a time-dependent oscillator, the derivative of
$\sigma$ coincides with the difference of kinetic and potential
energies. We show that this method can also be applied to both
fermionic oscillator and pair of the coupled constrained
oscillators. The last example is widely used in the theory of the
microwave background anisotropy. We have verified that there is no
problem with a selection of the vacuum state for the last system
in spite of some
discussions appearing in the literature \cite{gr,luk}.


\begin{thebibliography}{xxxx}
\bibitem{footnote1}{Under existence of global
vacuum we understand an existence of a state for which $\sigma$
has monotonic behavior at all range of $t$, whereas $\sigma$ for
the non-vacuum states oscillates.}
\bibitem{footnote2}{To be sure
that $\sigma=const$ minimizes the functional (\ref{func}) for some
family of the dependencies $\omega(t)$ including (\ref{onemore}),
it is sufficiently to write the extremum condition $\frac{\partial
Z(r,t)}{\partial r}=\frac{\partial}{\partial r}\int_{t_0}^T{\dot
\sigma}^2dt=2\int_{t_0}^T {\dot \sigma}\frac{\partial\dot
\sigma}{\partial r} dt$. To satisfy this condition, one can
suppose $\dot \sigma=0$. Substitution of the last equality into
Eq. (\ref{diffsig}) results in the particular family of
frequencies so that $ 2 \dot\omega -\frac{ \omega\, \ddot \omega
}{\dot\omega}=0$. }
\bibitem{footnote3}{Representation of the Dirac matrices is the same as in
Refs. \cite{lan4,cher}.}
\end{thebibliography}
\end{document}